\documentclass[twocolumn,showpacs,preprintnumbers,aps,amsmath,amssymb,prl]{revtex4}

\usepackage{graphicx}

\begin{document}

\title{Survival of the $d$-wave superconducting state near the edge of antiferromagnetism in the cuprate phase diagram}

\author{A. Hosseini}
\altaffiliation[Present address: ]{Materials Science and Engineering, Cornell University,
Ithaca, NY 14853.}
\author{D.M. Broun}
\altaffiliation[Permanent address: ]{Dept. of Physics, Simon Fraser University, Burnaby,
B.C., Canada, V5A 1S6.}
\author{D.E. Sheehy}
\altaffiliation[Present address: ]{Dept. of Physics, University of Colorado, Boulder, CO
80309}
\author{T.P. Davis, M. Franz, W.N. Hardy, Ruixing Liang and D.A. Bonn}
\affiliation{Department of Physics and Astronomy, University of British Columbia,
Vancouver, B.C., Canada  V6T 1Z1}

\date{\today}

\begin{abstract}


In the cuprate superconductor $\rm YBa_2Cu_3O_{6+x}$, hole doping in the $\rm CuO_2$
layers is controlled by both oxygen content and the degree of oxygen-ordering. At the
composition $\rm YBa_2Cu_3O_{6.35}$, the ordering can occur at room temperature, thereby
tuning the hole doping so that the superconducting critical temperature gradually rises
from zero to 20 K. Here we exploit this to study the $\hat{c}$-axis penetration depth as
a function of temperature and doping. The temperature dependence shows the $d$-wave
superconductor surviving to very low doping, with no sign of another ordered phase
interfering with the nodal quasiparticles. The only apparent doping dependence is a
smooth decline of superfluid density as $T_c$ decreases.

\end{abstract}


\maketitle




The high temperature superconductivity puzzle has important clues in the form of ordered
states of matter. One is the Mott insulator, where strong repulsion between electrons
fixes them to lattice sites, with spins ordered antiferromagnetically (AF). When doped
with holes, a superconductor is encountered, one in which paired holes condense into a
superfluid, but bear the stamp of the parent Mott insulator by exhibiting a $d$-wave
symmetry (dSC) favoured by strong electronic repulsion. This potential for finding new
states of matter drives much of the interest in systems with strong electronic
correlations and in the cuprates much of this focus lies on the border between the Mott
insulator and superconductor. In this regime of strong correlations and low hole-doping,
other exotic phases have been predicted
\cite{anderson,zhang,balents,vojta,varma,franz,chakravarty}, but experimentalists are
hampered by a lack of homogeneous samples in which doping spans this range. Here we
exploit a breakthrough \cite{liangI} in the growth of high purity crystals of $\rm
YBa_2Cu_3O_{6+x}$, where doping can be continuously tuned with room temperature
annealing. In the experiment presented here, the $\hat{c}$-axis penetration depth
$\lambda_c$ is measured on a sample in which critical temperature $T_c$ is tuned from
4-20 K. Throughout this range, the temperature dependence $\lambda_c(T)$ shows that the
nodal quasiparticles of the $d$-wave superconductor survive to very low doping, with no
sign of another ordered phase interfering with their behaviour. The change that does
occur with doping is a smooth decline of superfluid density as the $T_c$ is tuned down to
4 K.

A major advantage of working with the $\rm YBa_2Cu_3O_{6+x}$ system is that the
hole-doping can be reversibly changed by controlling oxygen in a CuO chain layer situated
0.42~nm away from the $\rm CuO_2$ planes that are the seat of the interesting physics.
When the chains are completely depleted of oxygen, at $\rm YBa_2Cu_3O_6$, the Mott
insulator is encountered and when they are filled $\rm YBa_2Cu_3O_7$ is a $d$-wave
superconductor with a critical temperature ($T_c$) near 90 K. Between these extremes,
Veal et al. \cite{veal} showed that the loosely-held dopant oxygens are mobile at room
temperature and their gradual ordering into chain structures pulls electrons from the
$\rm CuO_2$ planes \cite{zaanen}, increasing hole-doping and $T_c$ over time. An extreme
version of this effect occurs in a narrow window of oxygen content near $\rm
YBa_2Cu_3O_{6.35}$, where crystals quenched from over 570 $^o$C initially do not
superconduct, but become superconducting with a $T_c$ that increases with time if they
are allowed to order at room temperature. This opens the door to experiments over a wide
range of hole-doping, all on the same crystal, with no change in cation disorder.

Here we have exploited this in a measurement of the $\hat{c}$-axis penetration depth,
which provides information on the temperature and doping dependence of the superfluid
density. A key feature of the $d$-wave superconducting state is that the superfluid
density has a linear temperature dependence due to the presence of nodes in the $d$-wave
energy gap, in contrast to the exponential temperature dependence coming from the uniform
gap of a conventional superconductor. This linear term is seen in microwave measurements
of the magnetic penetration depth $\lambda_{ab}$ for superfluid screening currents
flowing in the $\rm CuO_2$ planes, which provide a direct measure of the superfluid
density \cite{hardy}.  A common measurement geometry involves a microwave magnetic field
applied parellel to the $\hat{ab}$-plane of a crystal that is thin in the
$\hat{c}$-direction, thus avoiding large demagnetizing effects. However, this geometry
includes an admixture of $\lambda_c$, the magnetic penetration depth for current-flow
perpendicular to the $\rm CuO_2$ planes, which becomes dominant when $\lambda_c$ becomes
extremely large at very low hole doping. So, here we turn our focus to measuring
$\lambda_c$ at low doping, since $1/\lambda_c^2$ is related to superfluid density,
although it is slightly less direct because it also depends on the coupling mechanism for
currents between the $\rm CuO_2$ planes.

Measurements of $\lambda_c$ were made in a 22.7~GHz cylindrical superconducting cavity
\cite{hosseiniprl} with the microwave magnetic field applied along the $\hat{a}$-axis of
a slab that was cut from a high purity crystal of $\rm YBa_2Cu_3O_{6+x}$ \cite{liangII}.
The sample was cut and polished to dimensions $\rm (a\times b\times c)=(1.803\times
0.203\times 0.391 mm^3$) that make contributions from $\lambda_{ab}$ negligible. The
temperature dependence of the $\hat{c}$-axis penetration depth $\Delta\lambda_c(T) =
\lambda_c(T) - \lambda_c(T_o)$ is extracted by measuring the shift in the cavity's
resonant frequency as the sample temperature is increased above a base temperature $T_o$
near 1.2 K. Importantly, $\lambda_c$ is large enough that the microwave fields penetrate
a substantial fraction of the crystal when it is in the superconducting state, but
completely penetrate the sample above $T_c$, making it possible to determine
$\lambda_c(T_o)\simeq \lambda_c(0)$ from the frequency shift seen when the sample is
warmed through $T_c$. A standard cavity perturbation relation $(\omega_o -
\omega)/\omega_o = \Gamma [1-2\tilde{\delta}/d\tanh(d/2\tilde{\delta})]$, is used to
extract the effective skin depth $\tilde{\delta}$ from the complex resonant frequencies
$\omega$ and $\omega_o$ of the cavity with and without the sample, respectively. $\Gamma$
is a geometrical constant obtained by an independent measurement on a Pb-Sn sample cut to
the same dimensions as the crystal and {\it d} is the thickness. The superfluid screening
length is extracted from the effective skin depth via $\lambda_c^{-2} = {\rm Re}\{
{\tilde{\delta}^{-2}}\} + \omega^2\epsilon_r/c^2$, where the term containing the
$\hat{c}$-axis dielectric constant $\epsilon_r$ is a small contribution that displacement
currents make to the screening. The value $\epsilon_r$ = 20 determined from far infrared
measurements \cite{homes} comes mainly from phonons and at a frequency of 22.7 GHz this
gives a displacement current contibution of 4.5$\rm \times 10^6 m^{-2}$, a correction of
order 1\% except for the very lowest $T_c$ samples and very near $T_c$. A weak
temperature dependence of $\Gamma$ due to demagnetizing effects has been corrected for.

\begin{figure}
\includegraphics[width=3.0in]{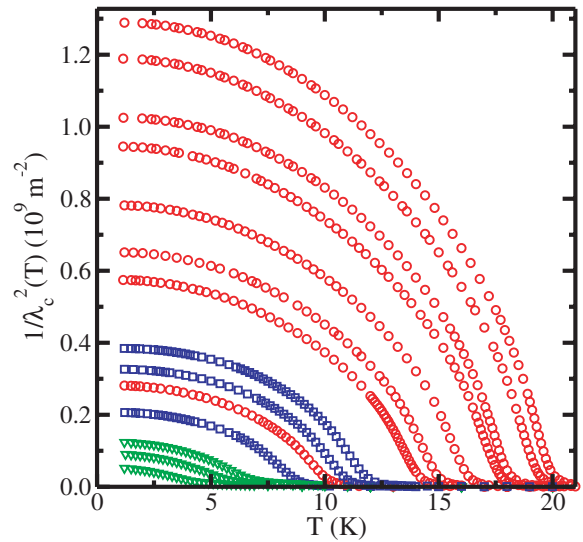}
\caption{\label{fig:superflu} (colour online) Growth of critical temperature and
$\hat{c}$-axis screening with increasing hole doping in $\rm YBa_2Cu_3O_{6+x}$,
x$\sim$0.35. $1/\lambda_c^2$ depends on the density of superconducting charge carriers
that participate in the screening of applied fields and has been determined by a cavity
perturbation measurement at microwave frequencies. The three sets of curves indicated by
squares, triangles and circles, correspond to 3 slightly different oxygen contents and at
each oxygen content the hole doping was allowed to drift up by way of gradual oxygen
ordering at room temperature.}
\end{figure}

\begin{figure}
\includegraphics[width=3.0in]{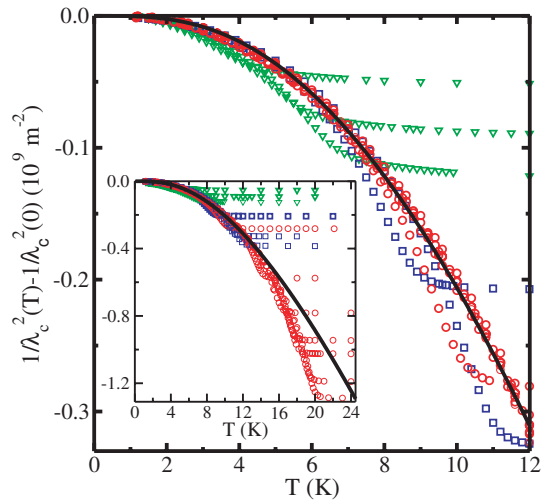}
\caption{\label{fig:superflu2}(colour online) A plot of $1/\lambda_c^2(T) -
1/\lambda_c^2(0)$ focusses attention on the depletion of the superfluid screening from
its low temperature value, reflecting the loss of superfluid density as quasiparticles
are thermally excited out of the superconducting condensate. The data for all of the
doping levels shown in Fig. \ref{fig:superflu} nearly collapse onto a single curve at low
$T$ with a power law close to $T^{2.5}$. The inset shows the same data over a wider
temperature range that spans the highest $T_c$. The solid curve is a fit to the low
temperature data of the higher $T_c$ samples, with a model that involves nodal
quasiparticles in a $d$-wave superconductor, scattered as they tunnel from one $\rm
CuO_2$ plane to the next.}
\end{figure}

Fig. \ref{fig:superflu} shows the rise of both $T_c$ and $1/\lambda_c^2$ as the hole
doping increases. The first remarkable feature of the data is the wide range of doping
that can be accessed in just one crystal, with $T_c$'s ranging from 4 to 20~K, all with
sharp transitions of width 1~K. The three sets of curves indicated by blue squares, green
triangles and red circles, show the results of three different stages in which the oxygen
content was adjusted in a narrow range near $O_{6.35}$ by annealing between 894 and 912
$^o$C in flowing oxygen, followed by a homogenization anneal at 570~$^o$C in a sealed
quartz ampoule with ceramic at the same oxygen content, then a quench to 0~$^o$C. At each
oxygen content the hole doping drifted up by way of gradual oxygen ordering at room
temperature, while remaining mounted in the microwave apparatus. A noteworthy feature of
the data set is that the results at slightly different oxygen contents overlap one
another over the course of the chain-ordering stage, indicating that for any particular
$T_c$, the behaviour of $1/\lambda_c^2(T)$ is dependent only on the hole-doping level,
not on the details of the degree of chain order and oxygen content. A second striking
feature is that at low temperatures the {\it absolute change} in $1/\lambda_c^2$ with
temperature has almost {\it no} doping dependence, in contrast to the large changes in
$1/\lambda_c^2(0)$ and $T_c$. Fig. \ref{fig:superflu2} focuses on this by plotting
$1/\lambda_c^2(T) - 1/\lambda_c^2(0)$, which reflects the depletion of the superfluid
density when quasiparticles are thermally excited out of the superconducting condensate.
There is a striking collapse of the data to a power-law temperature dependence at low
temperature, with very little doping dependence.

Qualitatively, the first conclusion that can be drawn from the power law behaviour is
that the nodes characteristic of a $d$-wave energy gap continue to govern the low
temperature properties. If a phase transition to another state with a gap opening at the
nodes had occurred, one would see the appearance of exponential rather than power-law
behaviour at low temperatures. The lack of such a change rules out transitions to other
exotic superconducting states such as $d+is$ or $d+id$, which break time-reversal
symmetry and open gaps at the nodes. The power law is close to $T^{2.5}$, the same as
that seen in measurements of $\lambda_c(T)$ in $\rm YBa_2Cu_3O_{6+x}$ at much higher hole
doping \cite{hosseiniprl}. The absolute magnitude of the $T^{2.5}$ term is remarkably
uniform, indicating that the nodal quasiparticles that control the low temperature
properties of the $d$-wave state are unaffected by any other physics developing as the
material approaches the limit of superconductivity in the phase diagram.

The $T^{2.5}$ power law for $1/\lambda_c^2(T)$ is different from the linear temperature
dependence seen in $\lambda_{ab}^2(T)$. This near-quadratic power law has been seen in
measurements of $\hat c$-axis penetration depth in $\rm La_{2-x}Sr_xCuO_{4+\delta}$
\cite{shibauchi}, $\rm Bi_2Sr_2CaCu_2O_8$ \cite{broun}, $\rm HgBa_2Ca_2Cu_3O_{8+\delta}$
\cite{panagopoulos}, as well as in $\rm YBa_2Cu_3O_{6+x}$ at higher oxygen dopings of x =
0.60, 0.95 and 0.99 \cite{bonnczech}. Understanding this difference from the
$\widehat{ab}$-plane behaviour requires a description of how holes tunnel between $\rm
CuO_2$ planes. When constructing such a model, it is important to note that if the
tunneling preserves the momentum of the hole and is also momentum-independent, then one
obtains for $1/\lambda_c^2(T)$ the linear $T$ dependence that is seen in
$1/\lambda_{ab}^2(T)$. Including scattering in the $\hat{c}$-axis tunneling is a natural
way to obtain a non-linear $T$-dependence, especially in light of the very high
$\hat{c}$-axis resistivity in these materials. The near-quadratic power law has been
modeled with an anisotropic scattering mechanism for $\hat{c}$-axis transport
\cite{radtke}, but recently Sheehy et al. have arrived at a detailed microscopic model
for such a process \cite{sheehy}. Within this model, the $T$-dependence depends crucially
on the nodal quasiparticles that are a defining feature of the $d$-wave superconducting
state. Written in terms of vector components relative to the nodes, these quasiparticles
have an energy spectrum given by $E_p = \sqrt{p_F^2v_F^2 + p_{\Delta}^2v_{\Delta}^2}$, a
Dirac spectrum where $p_F$ and $p_{\Delta}$ denote components of the quasiparticle
momentum perpendicular and parallel to the Fermi surface, $v_F$ is the Fermi velocity,
and the gap scale $v_{\Delta} = (\partial \Delta/\partial \phi)/p_F$ is the slope of the
$d$-wave energy gap $\Delta(\phi) = \Delta_o\cos{(2\phi)}$ as it increases with angle
$\phi$ away from the node. This Dirac spectrum takes the form of a very flattened cone,
since $v_F$ is known to be much greater than $v_{\Delta}$ \cite{lee}.

Scattering is introduced as random spatial variations in the tunneling matrix element. We
take the average, squared matrix element to be a Gaussian
\begin{equation}
\langle\vert t_{k-p}\vert^2\rangle = \frac{t_{\perp}^2}{\pi\Lambda^2}{\rm
exp}[-(k-p)^2/\Lambda^2]
\end{equation}
with overall magnitude $t_{\perp}^2$ and width $\Lambda$ that takes on small values so
that there is only a slight change in momentum when quasiparticles tunnel from plane to
plane. With this assumption, the $T$-dependent part of $1/\lambda_c^2$ is calculated
within standard BCS theory for a $d$-wave superconductor. The solid curve in Fig. 2 shows
that the model describes the temperature dependence at low temperatures extremely well,
with a single set of parameters for the scattering ($\hbar/\Lambda$ = 12 nm, $t_{\perp}$
= 26 meV) and parameters for the Dirac cone ($v_F$ = $\rm 3\times 10^5$ m/s,
$v_F/v_{\Delta}$ = 6.8). A doping-independent $v_F$ has been suggested by angle-resolved
photoemission (ARPES) measurements \cite{zhou}, so the doping-independent
$v_F/v_{\Delta}$ observed here might imply that $v_{\Delta}$ does not change
substantially through the low doping range. The key puzzle that has been solved is that
the nearly quadratic temperature dependence is crossover behaviour in the range
$v_{\Delta}\Lambda < k_BT < v_F\Lambda$. At lower temperatures a higher power law
dominates and at higher temperatures it is linear in $T$, but if $v_F/v_{\Delta}
>> 1$ there is a wide range where the temperature dependence is roughly quadratic \cite{sheehy}.
This model does not explain the much higher $T^5$ power law observed in the tetragonal
material $\rm HgBa_2CuO_{4+\delta}$ (which has been attributed to momentum-dependent
hopping) \cite{panagopoulos}, but it does explain the power law seen for $1/\lambda_c^2$
in the majority of orthorhombic cuprates \cite{bonnczech,shibauchi,panagopoulos,broun}.

\begin{figure}
\includegraphics[width=3.0in]{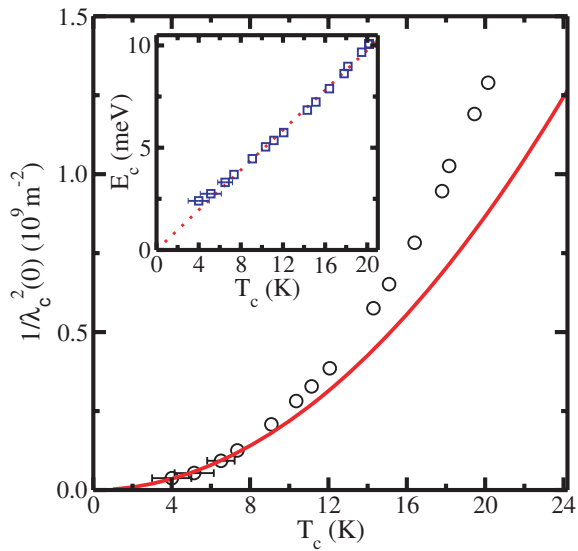}
\caption{\label{fig:correlation}(colour online) Relationship between $T_c$ and the
zero-temperature value of the $\hat{c}$-axis screening length.  The experimental
correlation between these two quantities (circles) can be explained with the model of
nodal quasiparticles cut-off by shrinking Fermi arcs. The inset shows the linear
relationship between the cut-off energy $E_c$ needed to model $1/\lambda_c^2$ and the
measured critical temperature $T_c$. This model produces a nearly quadratic dependence
between $1/\lambda_c^2$ and $T_c$, as indicated by the solid curve in the main figure.
The few degree discrepancy between the $T_c$ predicted by the model and the measured
$T_c$ is likely a consequence of critical fluctuations that govern the behaviour near
$T_c$.}
\end{figure}

This conventional treatment of the temperature dependence of $1/\lambda_c^2$ in terms
nodal quasiparticles in a $d$-wave superconductor completely fails to describe the doping
dependence of $1/\lambda_c^2$, which varies nearly quadratically with the sample's $T_c$,
as shown in Fig. \ref{fig:correlation}. Lee~and~Wen \cite{lee} have noted that the
superfluid density at higher doping displays a paradoxical disconnection between the
slope of the linear temperature dependence, whose doping dependence is thought to be weak
\cite{bonnczech,lemberger}, and the zero-temperature value, which declines rapidly as the
hole doping is reduced. The paradox lies in the fact that the zero-temperature value
seems to count the small number of holes added to the Mott insulator, but the temperature
dependence is consistent with a large $d$-wave gap on the large Fermi surface expected
for a high density of electrons. Here at much lower doping this phenomenology is
reflected in the doping-independent temperature dependence of Fig. \ref{fig:superflu2}
and the dramatic doping dependence of $1/\lambda_c^2(0)$ in Fig. \ref{fig:correlation}.
Following a suggestion by Ioffe and Millis \cite{ioffe}, this disconnection can be
modelled with an effective charge for the quasiparticles that falls to zero for states
away from the vicinity of the $d$-wave nodes, a view inspired by ARPES measurements that
show the Fermi surface restricted to small arcs of well-defined quasiparticles when
doping is decreased \cite{norman,yoshida}. The superfluid density then comes from a
truncated portion of Fermi surface \cite{lemberger} and Sheehy et al. \cite{sheehy} have
suggested that the doping dependence can be modeled with a Dirac cone that is cut off at
progressively lower energy $E_c$ as doping is decreased. The inset in Fig.
\ref{fig:correlation} displays a linearly falling $E_c$ that is extracted from
$1/\lambda_c^2(0)$ as doping and $T_c$ decrease, indicating a superfluid density that
steadily falls towards zero as $T_c$ does. What is remarkable is that in the face of this
depletion of superfluid density, the $d$-wave quasiparticle excitations nearest to the
nodes survive and continue to govern the temperature dependence. It has recently been
suggested that these nodal quasiparticles, seen here in samples with $T_c$ as low as 4~K,
even persist to dopings below the superconducting phase \cite{zhou} and show up in
thermal conductivity measurements in magnetic fields where $T_c$ is driven to zero
\cite{sutherland}. Thus the entire superconducting portion of the phase diagram, and
perhaps even more, is governed by nodal quasiparticles that survive even though the Fermi
surface is reduced to small arcs and the superfluid shrinks to almost nothing as the
magnetism is approached.

We thank S.A. Kivelson, P.A. Lee, S.-C. Zhang, I. Herbut, C.C. Homes and K.A. Moler for
discussions. This work was supported by the Natural Science and Engineering Research
Council of Canada and the Canadian Institute for Advanced Research.

\end{document}